\documentclass{hep99}
\usepackage{epsfig}
\newcommand{\xbj}{\mbox{$x_{Bj}~$}}
\newcommand{\xbjx}{\mbox{$x_{Bj}$}}

\newcommand{\Qsq}{\mbox{$Q^2~$}}

\newcommand{\et}{\mbox{$E_T~$}}

\newcommand{\as}{\mbox{$\alpha_s~$}}

\newcommand{\GeV}{\mbox{\rm ~GeV~}}

\newcommand{\GeVsq}{\mbox{${\rm ~GeV}^2~$}}
\newcommand{\GeVsqx}{\mbox{${\rm ~GeV}^2$}}

\newcommand{\ep}{\mbox{$e^{\pm}p~$}}

\newcommand{\als}{$\alpha_s$}

\def\lsim{\mathrel{\rlap{\lower4pt\hbox{\hskip1pt$\sim$}}
    \raise1pt\hbox{$<$}}}                
\def\gsim{\mathrel{\rlap{\lower4pt\hbox{\hskip1pt$\sim$}}
    \raise1pt\hbox{$>$}}}                

\begin{document}
\bibliographystyle{jetnotit}
\title{Determination of the Strong Coupling Constant in
Jet Production at HERA}
\author{\vspace{-0.2cm} Tancredi Carli$^1$}
%
%
\address{$^1$ Max Planck Institut f\"ur Physik\\
E-mail: {\tt Tancredi.Carli@desy.de} \\
On behalf of the H1 and ZEUS Collaborations. \\
Talk presented at the EPS conference 1999, Tampere, Finnland}
\abstract{
Recent $\alpha_s$ determinations from jet
production at HERA are presented. Three different observables
lead to consistent $\alpha_s$ values. The most accurate result is:
\begin{center}
$
\alpha_s(M_Z) =
0.1181 \pm 0.0030 \; ({\rm exp.}) 
^{+0.0039}_{-0.0046} \; ({\rm theo.}) 
^{+0.0036}_{-0.0015} \; ({\rm pdf})
$.
\end{center}
\vspace{-0.2cm}
} 
\maketitle
%
%
\begin{picture}(80,100)
\put(310,300){MPI-99-14}
\end{picture}
\vspace{-4.cm}
\section{Introduction \label{sec:intro}}
Perturbative Quantum Chromodynamics (pQCD), the theory describing short distance
strong interactions, has a single free parameter, the strong
coupling constant $\alpha_s$. This coupling depends on the renormalisation
scheme and the energy scale. At the reference scale, which
is customarily chosen as the mass of the $Z^0$ boson, the strong
coupling is presently known to an accuracy of about $4\%$~\cite{jet:bethke98}.
This is more than a factor of $10$ less precise than the coupling
of the weak force. 
Measurements of $\as$ over a wide energy range provide a fundamental
test of the evolution predicted by the QCD $\beta$-function and
can severely constrain physics beyond the Standard Model.

Precise determinations of $\alpha_s$~have~been~performed in $e^+ e^-$
annihilation processes over a large range of centre of mass
energies from $14 < \sqrt{s} < 189$ \GeV 
using jet rates, event shapes, $\tau$ decays and 
the inclusive ratio of hadron to lepton branchings~\cite{jet:bethke98}.

Jet production at hadron colliders is a particularly well suited 
process to measure the energy dependence of $\as$ in one single 
experiment. However, the presence of hadrons in the initial
state leads to the complication that the measured cross section
always depends on the product of $\alpha_s$ and 
the parton density functions (PDF). 
 

The high centre of mass energy ($s$) of HERA, colliding $27.5$ \GeV 
positrons on $820$ \GeV protons,
allows hard processes to be studied in deep-inelastic scattering (DIS).
Particularly well suited for pQCD studies is the Breit frame in which
the photon is purely space-like and collides head-on with the proton. 
Besides the photon virtuality, $Q^2$, the transverse energy $E_T$ 
produced in the Breit frame provides a hard scale. 

Jet cross sections at high $E_T$ are collinear 
and infrared safe observables which can be calculated in 
pQCD. 

\begin{figure}
\begin{picture}(80,100)
\put(-0,10){\epsfig{file=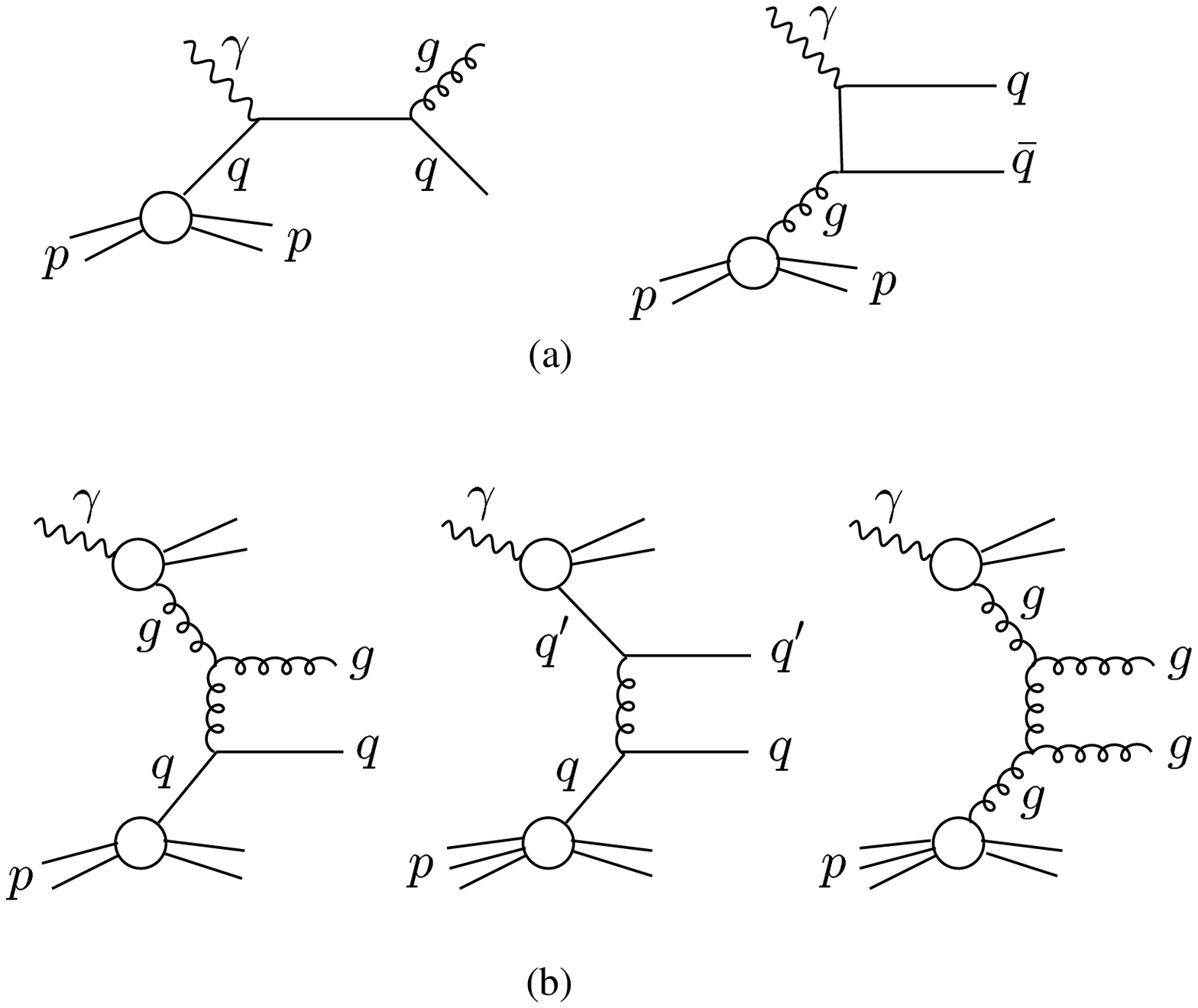,
 bbllx=274,bblly=451,bburx=480,bbury=602,clip,width=4.5cm}}
 \put(100,10){\epsfig{file=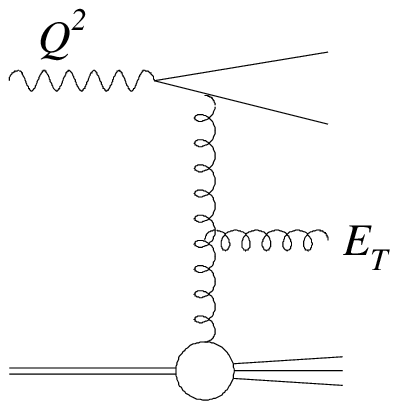,
 bbllx=31,bblly=69,bburx=173,bbury=203,clip,width=4.5cm}}
\end{picture}
\vspace{-0.9cm}
\caption{\it Examples of ${\cal O}(\alpha_s)$ and ${\cal O}(\alpha_s^2)$
Feynman diagrams dijet events 
in \ep-collisions.
\label{fig:feynjets}}
\vspace{-0.5cm}
\end{figure}

The jet cross section $\sigma_{jet}$ can be expanded to:
\begin{eqnarray}
\mbox{\vspace{-4.0cm}}
\sum_{a,n}
\alpha_s^n(\mu^2_{r}) 
\int_0^1 \frac{d\xi}{\xi} \,  
C_{a,n}(\xbjx/\xi,\mu^2_{r},\mu^2_{f},...)   
f_{a/p}(\xi,\mu_{f}^2), \nonumber 
\vspace{-0.2cm}
\label{eq:cross}
\end{eqnarray}
where $\xbjx = Q^2/(2 p \cdot \gamma)$, 
$\xi= \xbj  ( 1 + M^2_{j,j}/Q^2 )$ is in leading order (LO)
the fractional momentum of the struck parton in the proton
and $M_{j,j}$ the invariant dijet mass.
$\mu_{r}$ ($\mu_{f}$) is the renormalisation (factorisation)
scale. The perturbatively calculable coefficients $C_{a,n}$
are presently known to next-to-leading order (NLO).
The jet cross section is directly sensitive to 
$\alpha_s$ and the gluon ($f_{g/p}$) 
and quark ($f_{q/p}$) densities in the proton.
For low $Q^2$ (below $\approx 40 \GeVsqx$) jet cross sections 
calculated in NLO strongly depend on 
$\mu_r$~\cite{jet:carlinloshort,jet:mwhsshort}. 
pQCD analyses are therefore better performed at high $Q^2$,
where also hadronisation effects are smaller~\cite{jet:mwhsshort}.

Double differential inclusive dijet cross sections 
$d^2\sigma/d\xi dQ^2$ and
$d^2\sigma/dx dQ^2$ for $200 < Q^2 < 5000$ \GeVsqx, where jets
are defined by the inclusive $K_T$ 
algorithm~\cite{jet:inclkt} asking
for $\et > 5 \GeV$, $E_{T,1} + E_{T,2} > 17$ \GeV
and $-1 < \eta_{lab} < 2.5$~\fntext{1}{The pseudo-rapidity 
is defined by $\eta = \ln{\tan{\theta/2}}$.}
have been used for a preliminary extraction of the
gluon density in the range $0.01 \le \xi \le 0.1$ at fixed 
$\mu_f^2 = 200$ \GeVsqx~\cite{jet:h1dijetgluon98}. 
In this analysis $\as$ has been set to the world average.
$F_2$ data~\cite{jet:h1f298} for $200 < \Qsq < 650$ \GeVsq
have been included in a combined fit to get an additional
constraint on the quark densities.

At this conference new preliminary results have been presented
where the opposite strategy has been pursued. The parton 
densities have been taken from global fits to inclusive DIS
data and exclusive high-$E_T$ processes and then $\alpha_s$ is fitted.
Either sets of PDF with different input $\as$ have
been used to account for the anti-correlation between $\alpha_s$
and the PDF or the dependence on the PDF is included in the
systematic error. The final aim is to
perform a consistent simultaneous fit of both $\alpha_s$ and the
PDF.
%
\vspace{-0.5cm}
\section{\boldmath \as from Inclusive Jet Cross Sections \label{sec:h1}}
\vspace{-0.2cm}
The single inclusive jet cross section $d\sigma/dQ^2dE_T$
has been measured in four different $Q^2$ regions within
$ 150 < Q^2 < 5000$ \GeVsq and $0.2 < y = Q^2/(s \; \xbjx) < 0.6$
by the H1 collaboration. 
Jets are defined by the inclusive $K_T$ algorithm requiring
$\et > 5$ \GeV in the range $-1 < \eta_{lab} < 2.5$. 
The data are corrected for detector effects. 
Over the whole phase space reaching $E_T$ up to $50$ \GeV
a NLO calculation based on DISENT~\cite{jet:sey96d} 
using the CTEQ5M~\cite{th:cteq5} PDF
with $\mu_r^2 = E_T^2$ and $\mu_f^2 = 200$ \GeVsq 
describes the data. 
Hadronisation effects lower the NLO prediction by about 
$3 - 10 \%$~\cite{jet:mwhsshort}.
%
\begin{figure}
\epsfig{figure=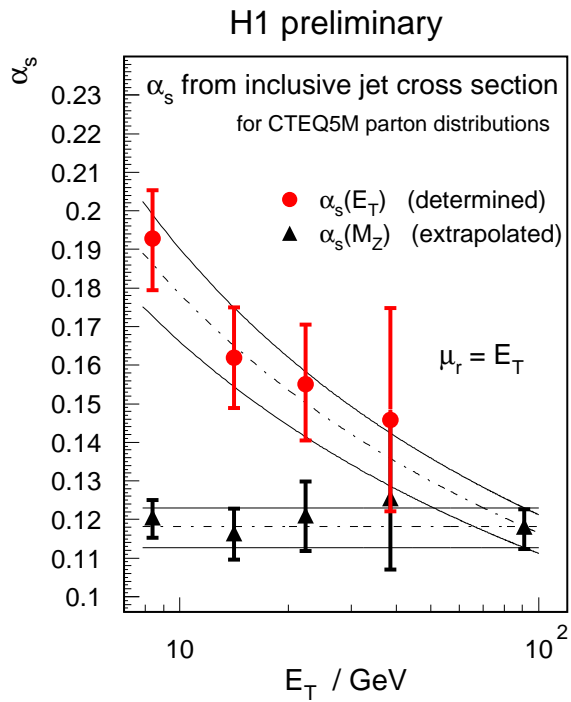,width=8cm}
\vspace{-0.9cm}
\caption{\it \as fit result versus the jet $E_T$.
The lines indicate the prediction of the
renormalisation group equation for the combined results
and their uncertainty. 
\vspace{-0.5cm}
\label{fig:h1alphas}}
\end{figure}
%
For each $E_T$ and $Q^2$ bin $\alpha_s$ is adjusted
to the measured cross section using a $\chi^2$ minimisation 
where the systematic errors are included in the 
fit~\cite{jet:wobischthesis}. All results are consistent
with each other. 
In Fig.~\ref{fig:h1alphas} the results are shown
as a function of $\et$ after having combined the four
$Q^2$ bins. The fitted $\alpha_s$ evolve as predicted by 
the renormalisation group equation. 
The combined $\alpha_s(M_Z^2)$ result is:
\begin{eqnarray}
\vspace{-0.1cm}
0.1181 \pm 0.0030 \; ({\rm exp.}) \, 
^{+0.0039}_{-0.0046} \; ({\rm theo.}) \, 
^{+0.0036}_{-0.0015} \; ({\rm pdf.}) \nonumber 
\vspace{-0.1cm}
\end{eqnarray}
The largest contribution to the experimental error comes from
the uncertainty on the hadronic energy scale. The theoretical
error is mainly determined by the uncertainty of the hadronisation
corrections and by the $\mu_r$ dependence.
The contribution obtained by varying $\mu_r^2$ by a factor of $4$ 
amounts to $^{+0.0025}_{-0.0034}$.
The hadronisation uncertainties are estimated in a very
conservative way using the QCD Monte Carlo models. 
Since the spread between this
models is rather small, half of the size of the
hadronisation corrections, but at least $3\%$, 
is assigned as systematic uncertainty.
If $\mu_r^2 = Q^2$ instead of $\mu_r^2 = E_T^2$ is used 
a consistent $\alpha_s(M_Z^2)$ result is obtained:
\begin{eqnarray}
\vspace{-0.1cm}
0.1221 \pm 0.0034 \; ({\rm exp.}) \,
^{+0.0054}_{-0.0059} \; ({\rm theo.}) \,
^{+0.0033}_{-0.0016} \; ({\rm pdf}) \nonumber 
\vspace{-0.1cm}
\end{eqnarray}
For~$\mu_r^2=\Qsq$~the~$\mu_r$~dependence~increases~by 
$\approx 40 \%$.

The systematic uncertainty on the PDF is difficult to estimate,
because none of the global analyses has yet provided PDF errors. 
Since the fit results of the different
groups, MRST~\cite{th:mrst} and CTEQ~\cite{th:cteq5}, 
are based on similar data sets 
and assumptions, the spread of different parametrisation does
not give a realistic error evaluation. Recently the CTEQ
collaboration~\cite{jet:cteqgluonerror} has investigated
possible variations on the gluon distribution using different
parametrisations and allowing the quality of the fit to be 
degraded. The resulting parametrisation is the first
step towards the estimation of uncertainties, but can not replace
a proper error analysis. 

Another point of concern is the influence of the $\alpha_s$ value 
used when deriving the PDF. If a strong correlation between
the fit result and the initial assumption on $\alpha_s$ was found, it would
be questionable whether the ansatz to fix the PDF and to extract
$\alpha_s$ would be meaningful.
This dependence can be tested by
means of series of parton distributions scanning from low to high
values of \als.
 
\begin{figure}
\epsfig{figure=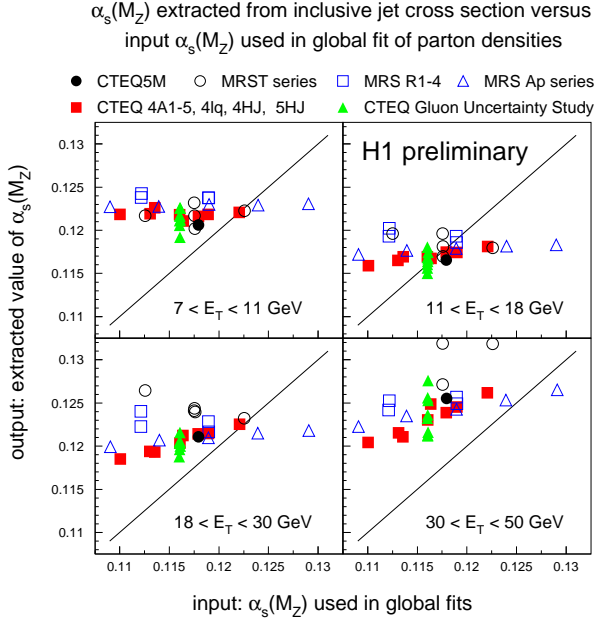,width=8cm}
\vspace{-0.8cm}
\caption{\it Dependence of the \as fit results on the
assumed PDF in four $E_T$ regions.
\label{fig:h1pdf}}
\vspace{-0.5cm}
\end{figure}

A comprehensive study of the fitted $\alpha_s$ on all available recent 
PDF, is presented in Fig.~\ref{fig:h1pdf}. The largest deviation
to the central result obtained with the CTEQ5M parametrisation
is quoted as systematic error. Only for the highest $E_T$ 
a slight dependence on the $\alpha_s(M_Z^2)$ value used in the global
fits is found. For all other bins no correlation is observed.
This is different from the observation made by the 
CDF collaboration~\cite{jet:cdfalphas} in an analysis of
single inclusive jet cross sections measured in 
$p \bar{p}$ collisions where a strong correlation
between the $\alpha_s$ used in the PDF and the fit result was found.

\section{\boldmath \as from Dijet Rates \label{sec:zeus}}
The ZEUS collaboration has measured dijet cross sections and rates 
in the phase space $470 < \Qsq < 20000$ \GeVsq and $0 < y < 1$.
Jets are defined as in the dijet analysis described before,
except that the two highest $E_T$ (and not all) jets have to fulfil
$-1 < \eta_{lab} < 2$. In addition, only events with
exactly two jets are considered. The exclusive
dijet cross sections as a function of $E_T$, $M_{j,j}$, 
$\eta_{lab}$, $\xbj$, 
$z_{p,1} = E_{1} \, (1 - \cos{\theta_i}) /
(\sum_{k=1,2} E_k ( 1 - \cos{\theta_k}))$, 
and $\xi$ 
are - both in shape and in absolute magnitude - 
well described by a NLO calculation using
CTEQ4M~\cite{th:cteq4} for $\mu_r^2 = \mu_f^2 = Q^2$. 
The data are corrected for detector, electroweak radiation
and hadronisation effects.
Hadronisations corrections are at most $10\%$ and decrease with
increasing $Q^2$. For $Q^2 > 5000 \GeVsq$ the contributions
from $Z^0$ exchange become visible. They have been subtracted
from the data. 

\begin{figure}
\epsfig{figure=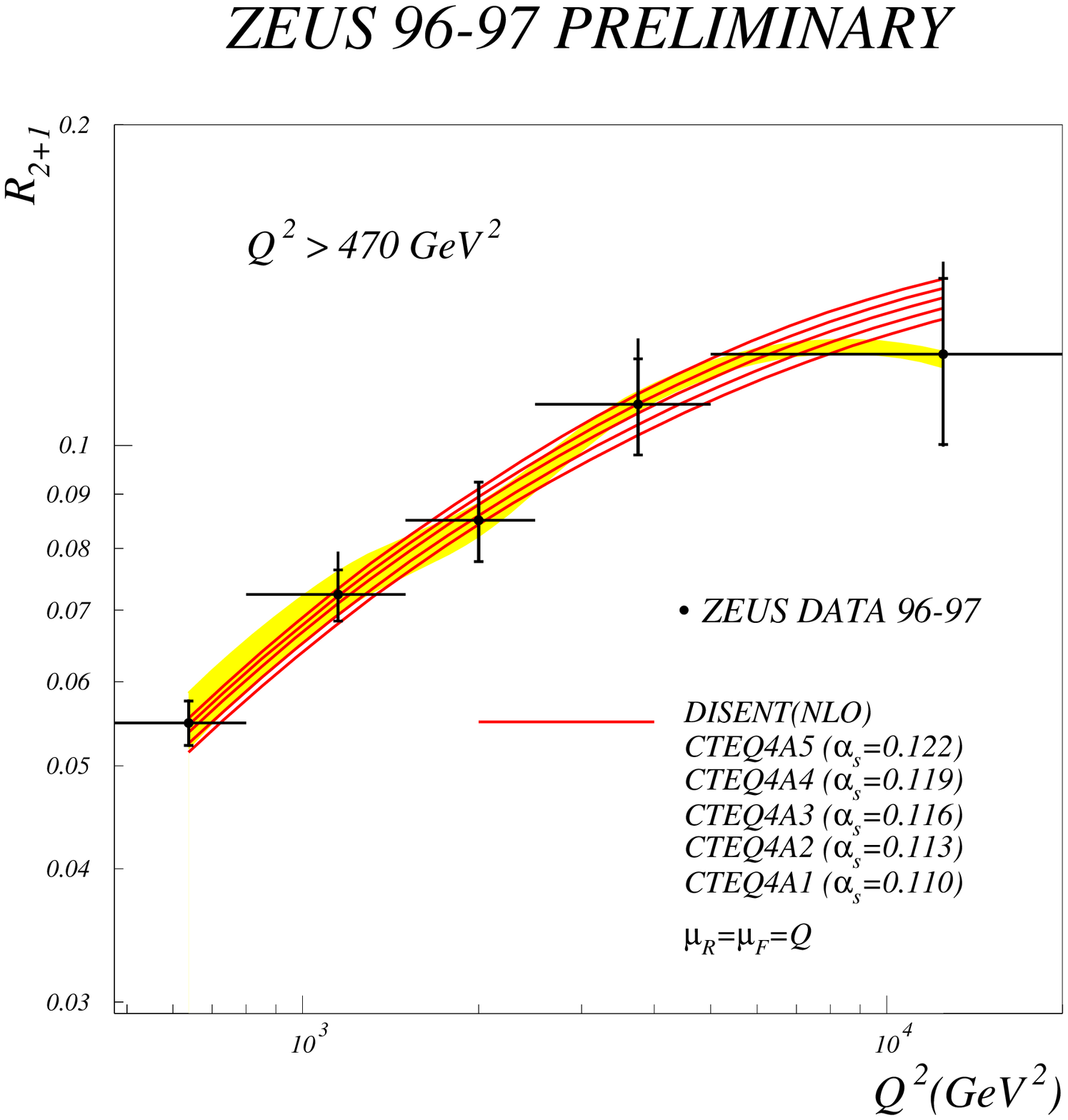,width=8cm}
\vspace{-0.8cm}
\caption{\it Exclusive dijet rate as a function of $Q^2$.
\label{fig:zeusr2}}
\vspace{-0.5cm}
\end{figure}
The good description of the data  allows an extraction
of $\alpha_s$ from  
$R_{2+1}(Q^2) =\sigma_{2+1}(Q^2)/\sigma_{\rm tot}(Q^2)$.
The dijet rates shown in Fig.~\ref{fig:zeusr2} are compared
to NLO QCD calculations based on
the CTEQ4A PDF series which have been fitted for different
$\alpha_s$ assumptions. This series is also used to
parametrise the anti-correlation between $\alpha_s$ and the 
PDF~\cite{tassithesis}.   
The value of $\alpha_s$ is obtained from a $\chi^2$ minimisation:
\begin{eqnarray}
\vspace{-0.1cm}
0.120 \pm 0.003 \; ({\rm stat.}) \, 
^{+0.005}_{-0.006} \; ({\rm exp.}) \, 
^{+0.003}_{-0.002} \; ({\rm theo.}) \nonumber 
\vspace{-0.1cm}
\end{eqnarray}
The experimental systematic error is dominated by the uncertainty
on the hadronic energy scale ($\pm 0.005$). QCD Model dependencies 
to unfold the data to parton level account for an error of
$^{+0.001}_{-0.002}$. The theoretical uncertainty is estimated
by varying the renormalisation scale $\mu_r^2 = Q^2$ by a factor of
$2$ ($\pm 0.002$), 
by replacing the assumed PDF by MRST sets ($-0.002$) and by
taking DISASTER++~\cite{jet:graudenz97a} instead of DISENT
($\pm 0.002$). If $\mu_r^2$ is varied by a factor of $4$ as in 
the previous analysis, the $\mu_r$ uncertainty increases
to $\pm 0.004$.
The dijet rates change only by at most $2.5\%$ when the PDF from 
the CTEQ gluon uncertainty study is used. The influence on the
$\alpha_s$ result is therefore expected to be small.
 
%
\vspace{-0.1cm}
\section{\boldmath $\alpha_s$ from Differential Jet Rates \label{sec:y2}}
H1 has measured $1/N_{tot} \, (dn/dQ^2dy_2)$ 
for $Q^2 > 150$~\GeVsq and $0.1 < y < 0.7$. 
All particles are clustered on the basis of a 
distance measure $y = d_{i,j}/{\rm scale}$, where  
$d_{i,j} = 2 \min{(E^2_i,E^2_j)} ( 1 - \cos{\theta_{i,j}})$,
until $2+1$ jets are found~\fntext{2}{$+1$ denotes the proton remnant.
$E_i$ is the energy of the particles 
and $\theta_{i,j}$ the angle between them.}.  
$y_2$ is the smallest value of $y$ where two jets can be resolved. 
It is therefore a measure for the jet structure.
Two jet algorithms are used.
In both cases the data are described by a NLO
calculation based on CTEQ5M~\cite{tobienthesis}.   

The $K_T$ jet algorithm for DIS~\cite{jet:ktalogodis} is
applied in the Breit frame and ${\rm scale}$ is set to $100$~\GeVsqx.
The proton remnant $p$ is included 
as particle with $d_{i,p} = 2 E^2_i ( 1 - \cos{\theta_i})$.
After a cut $y_2 > 0.8$ only events with $E_T > 10$~\GeV
in the Breit frame remain. 
A $\chi^2$ minimisation gives $\alpha_s(M_Z^2)$:
\begin{eqnarray}
\vspace{-0.1cm}
0.1143 \; ^{+0.0075}_{-0.0089} \; ({\rm exp.}) \,
^{+0.0074}_{-0.0064} \; ({\rm theo.}) \,
^{+0.0008}_{-0.0054} \; ({\rm pdf}) \; \; \; \; \;  \nonumber 
\vspace{-0.1cm}
\end{eqnarray}

A variant of the JADE algorithm is applied in the HERA laboratory
frame and the squared invariant mass of the hadronic final state $W^2$ 
is used as scale, applied in the HERA laboratory 
(``mod. Durham algorithm''). 
A massless particle is added in the clustering procedure to account
for the escaped longitudinal momentum. 
$y_2$ is required to be above 
$0.005$ to select hard processes. The fit result is:
\begin{eqnarray}
\vspace{-0.1cm}
\nonumber  
0.1189 \; ^{+0.0064}_{-0.0081} \; ({\rm exp.}) \,
^{+0.0059}_{-0.0046} \; ({\rm theo.}) \,
^{+0.0013}_{-0.0055} \; ({\rm pdf}) \; \; \; \; \;  \nonumber 
\vspace{-0.1cm}
\end{eqnarray}

The quoted values are obtained for $Q^2 > 575$ \GeVsqx,
the lower $Q^2$ regions give consistent results.
The experimental errors are dominated by the energy scale
uncertainty and by the QCD model dependence of the detector
correction. The theoretical uncertainty is given by the hadronisation
correction ($\pm 0.0026$) and the $\mu_r$ dependence:
$^{+0.0053}_{-0.0038}$ for $\mu_r^2 = Q^2$ and 
$^{+0.0025}_{-0.0007}$ for $\mu_r^2 = E_T^2$.
The correlation between the fitted $\alpha_s$
and the value assumed in the PDF seems to be - in particular
at low $Q^2$ - stronger for jet
rates than for jet cross sections. 

\section{Summary \label{sec:sum}}
\begin{figure}
\epsfig{figure=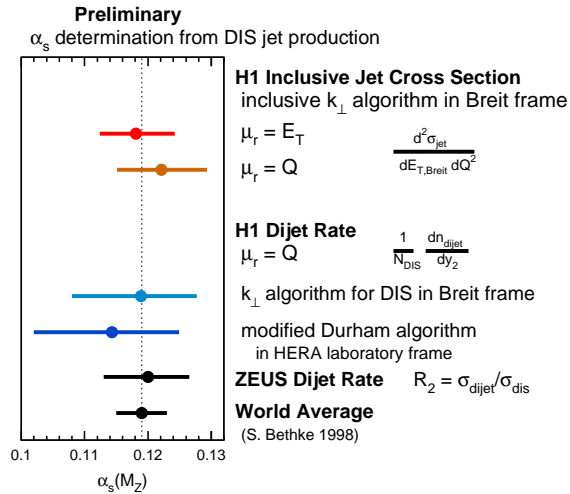,width=8cm}
\vspace{-0.8cm}
\caption{\it Summary of recent \as results from jets. 
\label{fig:alphsum}}
\end{figure}

$\alpha_s(M_Z)$ has been determined in NLO from
jet observables in DIS at HERA (see Fig.~\ref{fig:alphsum}).
The result is consistent with the world average 
and has an error of about $0.006$.
This is quite a remarkable result, since it is
only slightly less precise than the
world average value having an error of $0.004$~\cite{jet:bethke98}.

In future, fitting technique using $F_2$ and jet data
should be further developed to allow a simultaneous fit
of $\alpha_s$ and the PDF from HERA data alone. 

%
\section{Acknowledgement}
I would like to thank my colleagues P. Schleper,
S. Schlenstedt, E. Tassi and M. Weber 
for the critical reading of the manuscript.
%

\end{document}